\documentclass[%
reprint,
superscriptaddress,
amsmath,amssymb,
aps,
prx,
]{revtex4-2}

\usepackage{chemformula} 
\usepackage[T1]{fontenc} 
\usepackage{amsmath} 

\begin{document}

	\title{Non-Markovian non-equilibrium modeling of experimental cell-motion trajectories 
	reveals dependence of propulsion-force correlations on solvent viscosity}
	
	
	\author{Anton Klimek}
	\affiliation{Fachbereich Physik, Freie Universit{\"a}t Berlin, 14195 Berlin, Germany}
	
	\author{Prince V. Baruah}
	\affiliation{Department of Physics, Indian Institute of Science, 560012 Bangalore, India}
	
	\author{Prerna Sharma}
	\affiliation{Department of Physics, Indian Institute of Science, 560012 Bangalore, India} 
	\affiliation{Department of Bioengineering, Indian Institute of Science, 560012 Bangalore, India}
	
	\author{Roland R. Netz}
	\affiliation{Fachbereich Physik, Freie Universit{\"a}t Berlin, 14195 Berlin, Germany}
	\email{rnetz@physik.fu-berlin.de}

	\begin{abstract}
        Cell motility underlies many biological processes, including cancer metastasis, bacterial infection, and evolutionary adaptation.
		We introduce a non-equilibrium single-cell motility model inspired by the generalized Langevin equation, which accounts for hydrodynamic friction 
		and correlated propulsion force.
        From video microscopy of Chlamydomonas reinhardtii algae and Salmonella typhimurium bacteria we extract the propulsion-force dynamics on the single-cell level, which we find to exhibit multi-exponential correlations, not captured
        by literature non-equilibrium cell-motility models.
        Based on our data-driven model,
        we predict the effective cell diffusivities beyond experimentally resolved timescales and demonstrate a diffusivity maximum at intermediate solvent viscosity for both cell types. 
        This means that cells adapt their propulsion-force characteristics according to the solvent viscosity.
        In addition, our model predicts the power output of single cells, which is on the order of $\rm{aW}$ for the salmonella and $\rm{fW}$ for the algae.

	\end{abstract}
	
	\maketitle
	
	\section{Introduction}
    The motion of individual cells often influences their biological functioning and is a key factor determining the risk of infection by pathogens \cite{josenhans2002role,zegadlo2023bacterial}, the probability of cancer metastasis \cite{jang2025kindlin,chen2025mechanisms} and the reaction of cell populations to environmental changes \cite{levien2021non}, just to name a few examples.
    Actively moving cells convert chemical energy into locomotion, which renders the process non-equilibrium \cite{mizuno2007nonequilibrium}.
    Modeling non-equilibrium processes using first-principles and starting from a molecular level is notoriously difficult \cite{zwanzig2001nonequilibrium,klages2007microscopic,prigogine2017non,netz2024derivation}.
    Thus, many modeling approaches use a coarse-grained level to describe the motion of single cells.
    Widely used models include the active Ornstein-Uhlenbeck (OU) process \cite{martin2021statistical,jones2021stochastic}, active Brownian motion \cite{sprenger2020active,tsekov2013brownian} and cell-specific models that are based on structural and dynamical properties of the specific cell \cite{zhang2023run,heyn2025cell,sadhu2023minimal,callan2016actin}.
    Cell motion models are often chosen to describe or predict a certain quantity of interest.
    For instance, cell-specific models usually aim to explain the relation between cell-internal processes and the observed motion.
    In contrast, statistical mechanics models focus on the quantification of motility patterns to draw conclusions about dynamical properties that can influence biological function.
    For example, the cell speed and diffusivity of bacterial cells can determine how long they survive in the gut \cite{wiles2020swimming} or how infectious they are \cite{josenhans2002role,zegadlo2023bacterial}.

    The diffusivity describes the long-time space coverage of passively moving particles and depends on the temperature and viscosity of the surrounding medium, but can also be used to characterize active motion, which then also depends on the activity of the cell.
    The diffusivity is generally given by the slope of the mean squared displacement (MSD) at large times, where it depends linearly on time.
    The MSD is a key quantity that describes how far a cell moves in average over a time interval $t$ and is defined as
    \begin{equation}
    \label{eq_msd_def}
        C_{\rm{MSD}}(t) = \langle (x(t)-x(0))^2 \rangle\,,
    \end{equation}
    where $\langle\cdot\rangle$ denotes a time average and $x(t)$ is the position of the cell at time $t$.
    The typical MSD of an actively swimming cell exhibits a ballistic regime $C_{\rm{MSD}}(t)\propto t^2$ at short times, which represents directed motion, and a diffusive regime at long times $C_{\rm{MSD}}(t) = 2\mathcal{D}Dt$, which is determined by a finite persistence and reorientation effects \cite{klimek_data-driven_2024}.
    Here, $D$ denotes the effective diffusivity and $\mathcal{D}$ the spacial dimension.
    The MSD crossover between the short-time ballistic regime and the long-time diffusion is cell specific and can exhibit oscillatory or power-law behavior \cite{klimek_2024_cell,dieterich2008anomalous}.

    The long-time diffusivity $D$ of a cell determines how fast the cell samples its environment over long periods of time, which is highly relevant to determine the potency of pathogens and other biological factors related to how fast cells explore space.
    On the other hand, short time dynamics strongly influence cell-to-cell interactions, which lead to the emergence of collective effects \cite{martin2021statistical,caprini2021collective,sokolov2012physical,sathe2016cellular}, that are widely studied and relevant for cells that live in dense suspensions \cite{caprini2021collective,sathe2016cellular}.
    Even though short- and long-time effects are crucial to develop a complete description of cell motion, it is experimentally challenging to obtain trajectories that resolve both.

    The most general equation of motion that can be derived from the many-body Hamiltonian for a cell in an isotropic and translationally invariant environment is the generalized Langevin equation (GLE) of the form \cite{zwanzig2001nonequilibrium,dalton2024memory,netz2024derivation,vroylandt2022derivation}
	\begin{equation}
		\label{eq_gle}
		\ddot{x}(t) = -\int_{t_0}^{t} \Gamma_v(t-t') \dot{x}(t') dt' + F_R(t) \,,
	\end{equation}
    where $x(t)$ is the position along one cartesian coordinate.
    Here, $\ddot{x}(t)$ denotes the acceleration, $\dot{x}(t)$ the velocity, $\Gamma_v(t)$ is the velocity friction kernel, describing the viscous coupling to the environment, and $F_R(t)$ is the effective propulsion force that pushes the cell forward.
    We denote the second moment of the propulsion force by
    \begin{equation}
		\label{eq_fdt_neq}
		\langle F_R(t) F_R(0) \rangle = \Gamma_R (t) \,.
	\end{equation}
    It should be noted that cells moving at low Reynolds number actually create a force dipole and $F_R(t)$ is only the net-resulting force that pushes the cell against the friction of the environment \cite{zhu2013low,winkler2016low,mondal2021strong}.

    From a single measurement of a trajectory, one cannot deduce both, $\Gamma_R(t)$ and $\Gamma_v(t)$ \cite{netz2025time,mitterwallner2020non}.
    This would require either an independent measurement, for instance measuring $\Gamma_v(t)$ by dragging a cell at constant speed through the solvent, or knowledge about the structure of either $\Gamma_R(t)$ or $\Gamma_v(t)$ \cite{netz2025time,lucente2023revealing,lucente2023statistical}.
    In previous work we circumvented this issue by modeling the motion of various cell types by an effective GLE model with $\Gamma^{\rm{eff}}(t)=\Gamma_R^{\rm{eff}}(t)=\Gamma_v^{\rm{eff}}(t)$ \cite{mitterwallner2020non}, which enabled us to classify single cells by their motion \cite{klimek_data-driven_2024} and to determine cell-to-cell heterogeneity in noisy cell-motility data sets \cite{klimek_2024_cell}.
    However, the effective kernel approach using $\Gamma^{\rm{eff}}(t)$ does not yield information on the non-equilibrium character of the motion, and it cannot be applied to cells with non-Gaussian statistics.
    In particular, this means that the effective description cannot be used to infer collective effects of cells that interact with each other, which typically renders the statistics non-Gaussian.

    In this paper, we introduce an alternative approach: By using a hydrodynamic prediction for $\Gamma_v(t)$, we extract the actual non-equilibrium propulsion-force correlation $\Gamma_R(t)$ from video microscopy data of individual \textit{chlamydomonas reinhardtii} algal cells \cite{jeanneret2016brief,harris2001chlamydomonas} as well as for individual \textit{salmonella typhimurium} bacterial cells \cite{galan2021salmonella,stecher2008motility}.
    We resolve the short-time behavior on a scale of $0.002\,\rm{s}$.
    Our data-driven non-equilibrium model enables us to predict the long-time diffusivity $D$, which describes the cell motility on a scale of tens to hundreds of seconds, much longer than the experimental observation time.
    Further, we vary the surrounding viscosity of the cells by adding increasing amounts of PEG in separate measurements and find a maximum in the diffusivity at intermediate viscosity for both cell types.
    This implies that the salmonella and algal cells adapt their propulsion-force characteristics to the viscosity of their environment.

    \section{Results \& Discussion}
    \begin{figure*}
		\includegraphics{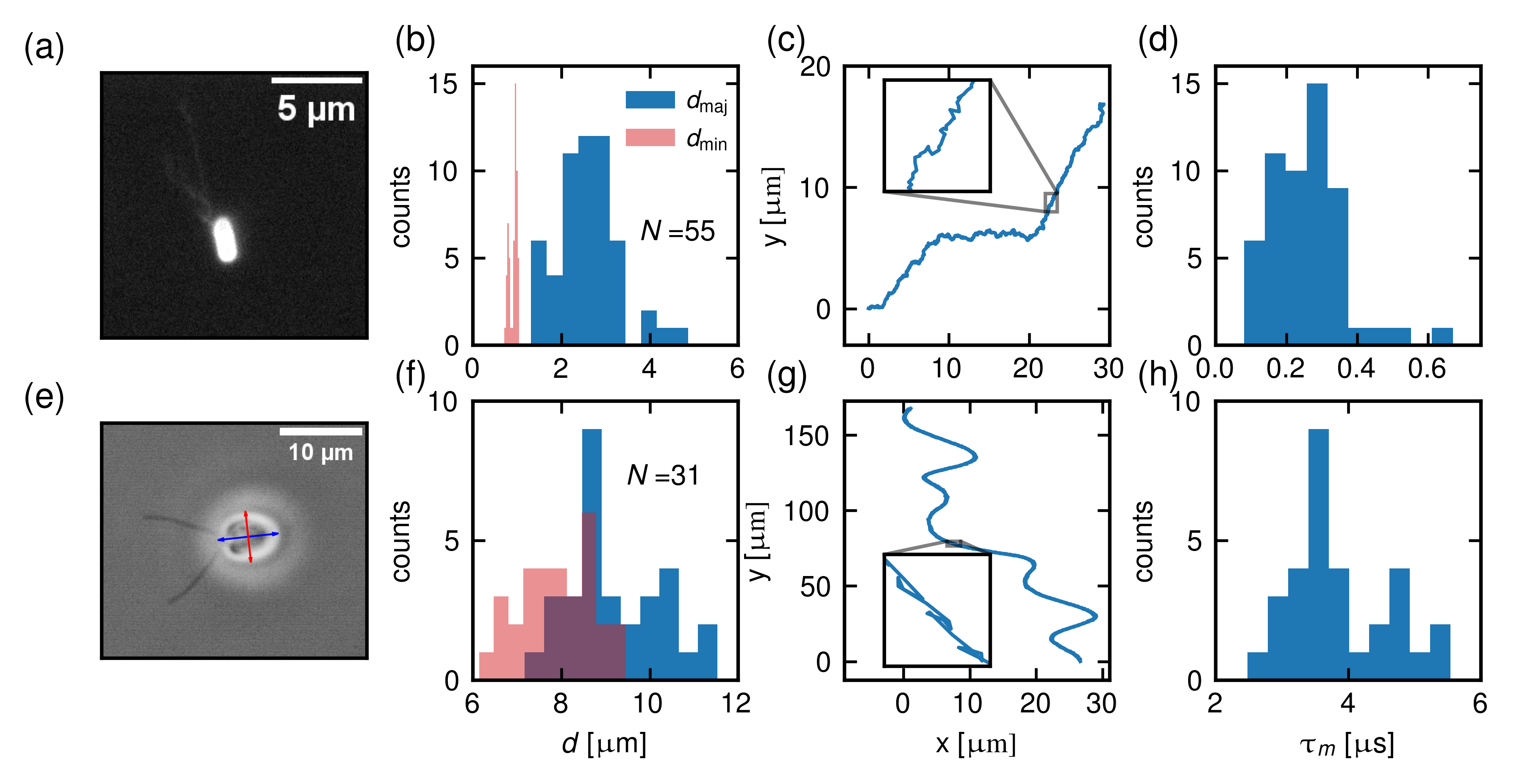}
		\caption{Microscopy image of (a) a single salmonella (e) a single algal cell, where the major axis is indicated by a blue line and the minor axis by a red line.
        The corresponding distributions of cell diameters measured in buffer solution of viscosity $\eta=0.89\,\rm{mPas}$ for (b) salmonella, (f) algae.
        Exemplary trajectory of a single (c) salmonella cell, (d) algal cell,  extracted from video microscopy, where the insets show enlarged details of the respective trajectory.
        The total trajectory length is $2.09\,\rm{s}$ for the salmonella and $2.272\,\rm{s}$ for the algal cell (see SI for trajectory-length distributions).
        Distribution of inertial times $\tau_m$ for (d) salmonella, (h) algae. 
		}
		\label{fig1}
	\end{figure*}
    Exemplary microscopy images of a single salmonella and a single algal cell are shown in Fig.~\ref{fig1}a,e, which showcase the ellipsoidal shape of the respective cell bodies and from which the cell position is determined as the geometric center.
    Details about the cell culturing and imaging are given in the SI.
    The salmonella exhibit lengths around $d_{\rm{maj}}\approx3\,\mu\rm{m}$ and widths around $d_{\rm{min}}\approx1\,\mu\rm{m}$ as shown in Fig.~\ref{fig1}b, whereas the algae are less elongated with $d_{\rm{maj}}\approx9\,\mu\rm{m}$ and $d_{\rm{min}}\approx8\,\mu\rm{m}$, as shown in Fig.~\ref{fig1}f.
    Both cell types move mainly in the direction of the major axis, which is the direction with the smallest hydrodynamic friction.
    While the salmonella bundle their flagella together on one side and push themselves in the opposite direction by rotating the flagella \cite{minamino2018salmonella,alam1984flagella}, the algae pull themselves in the direction of the flagella-anchoring with a breast-stroke-like motion \cite{jeanneret2016brief,klimek_data-driven_2024}.
    To record long trajectories, we confine the cells between two non-sticky surfaces with a distance of $5\,\mu\rm{m}$ for the salmonella (between coated glass slide and PDMS) and $30\,\mu\rm{m}$ for the algae (between two coated glass slides).
    See SI for experimental details.
    This setup ensures that the cells move in the microscope focus for longer times, which we set to the middle between the glass plates.
    The respective distance of the glass slides is chosen such that cells rarely move out of focus and contact friction with the confining walls is absent when cells move in the focal plane.
    Such confinement reflects the natural soil habitat of algal cells \cite{jeanneret2016brief,mondal2021strong}, whereas salmonella typically occupy environments with different degrees of confinement \cite{galan2021salmonella}, potentially resulting in motility behaviors distinct from those observed here.
    We record trajectories in a dilute cell suspension, such that cell-to-cell interactions are negligible.
    The experimental setup yields two-dimensional trajectories and we model the motion in the $y$-direction of the lab frame by the same GLE Eq.~\eqref{eq_gle} as for the $x$-direction.
    Thus, all correlation functions shown are averaged over the two spacial dimensions.
    We mention in passing that in certain scenarios of interacting organisms, modeling approaches in a comoving frame are more appropriate, especially in the case of non-homogeneous environments \cite{azuara2025modelling}.
    Exemplary trajectories for a salmonella and an algal cell are shown in Fig.~\ref{fig1}c,g respectively.
    By inverting the GLE Eq.~\eqref{eq_gle}, we can extract the propulsion force for a given trajectory, if $\Gamma_v(t)$ is known.
    The propulsion force for a discrete trajectory then reads
    \begin{equation}
    	\label{eq_random_force_extraction}
    	F_R^i = \ddot{x}^i + \Delta \sum_{j=1}^{i-1} \Gamma_v^{i-j} \dot{x}^j + \frac{\Delta}{2} (\Gamma_v^0 \dot{x}^i + \Gamma_v^i\dot{x}^0)\,,
    \end{equation}
    where time-dependent functions at discrete time points are denoted by $f(t)=f(i\Delta)=f^i$ with time steps $\Delta$ and the acceleration and velocity are determined from trajectories via discrete derivatives.
    
    We observe mean cell speeds of $25\,\mu\rm{m}/\rm{s}$ for the salmonella and $166\,\mu\rm{m}/\rm{s}$ for the algae in aqueous solutions of viscosity $\eta=0.89\,\rm{mPas}$, which is in accordance to literature values \cite{laszlo1981aerotaxis,liu2020swimming,jeanneret2016brief}.
    The speeds at which these organisms move and their elliptical shapes imply that their friction is frequency independent on the time scales of observation according to theoretical predictions for spherical objects in water \cite{kim2013microhydrodynamics,kiefer2025effect}.
    This suggests that the friction kernel in Eq.~\eqref{eq_gle} is given by
    \begin{equation}
        \label{eq_delta_kern}
        \Gamma_v(t) = 2 \delta (t)/ \tau_m\,,
    \end{equation}
    where the inertial time $\tau_m = m/\gamma$ is defined by the mass of the cell $m$ and the friction coefficient with the surrounding liquid $\gamma$.
    Therefore, $\Gamma_v(t)$ takes the discrete values $\Gamma_v^0=2/(\tau_m\Delta)$ and $\Gamma_v^i=0$ for $i>0$.
    We estimate $m$ by the cell volume and approximate the density as the density of water, which leads to a mass comparable to direct measurement of dried suspensions \cite{chioccioli2014flow}.
    Further, the hydrodynamic friction of an ellipsoid in confined laminar Stokes flow is given by
    \begin{equation}
    \label{eq_friction_hydro}
        \gamma = 3\pi \eta d_{\rm{maj}} f_{\rm{w}} f_{\rm{e}}\,,
    \end{equation}
    where $f_{\rm{e}}$ is a factor that accounts for the ellipsoidal shape of the cell body \cite{kim2013microhydrodynamics,zendehroud2024molecular} and $f_{\rm{w}}$ accounts for the hydrodynamic interaction with the confining glass walls \cite{ganatos1980strong}.
    Details on $f_{\rm{e}}$ and $f_{\rm{w}}$ are given in the SI.
    We find that the inertial times $\tau_m$ for individual cells are in the order of tenths of microseconds for the salmonella and microseconds for the algae, as shown in Fig.~\ref{fig1}d,h.
    This implies that the cell motion is overdamped and inertial effects are negligible on all relevant time scales, the smallest of which is the recording time step $\Delta=0.002\,\rm{s}$.
    \begin{figure*}
		\includegraphics{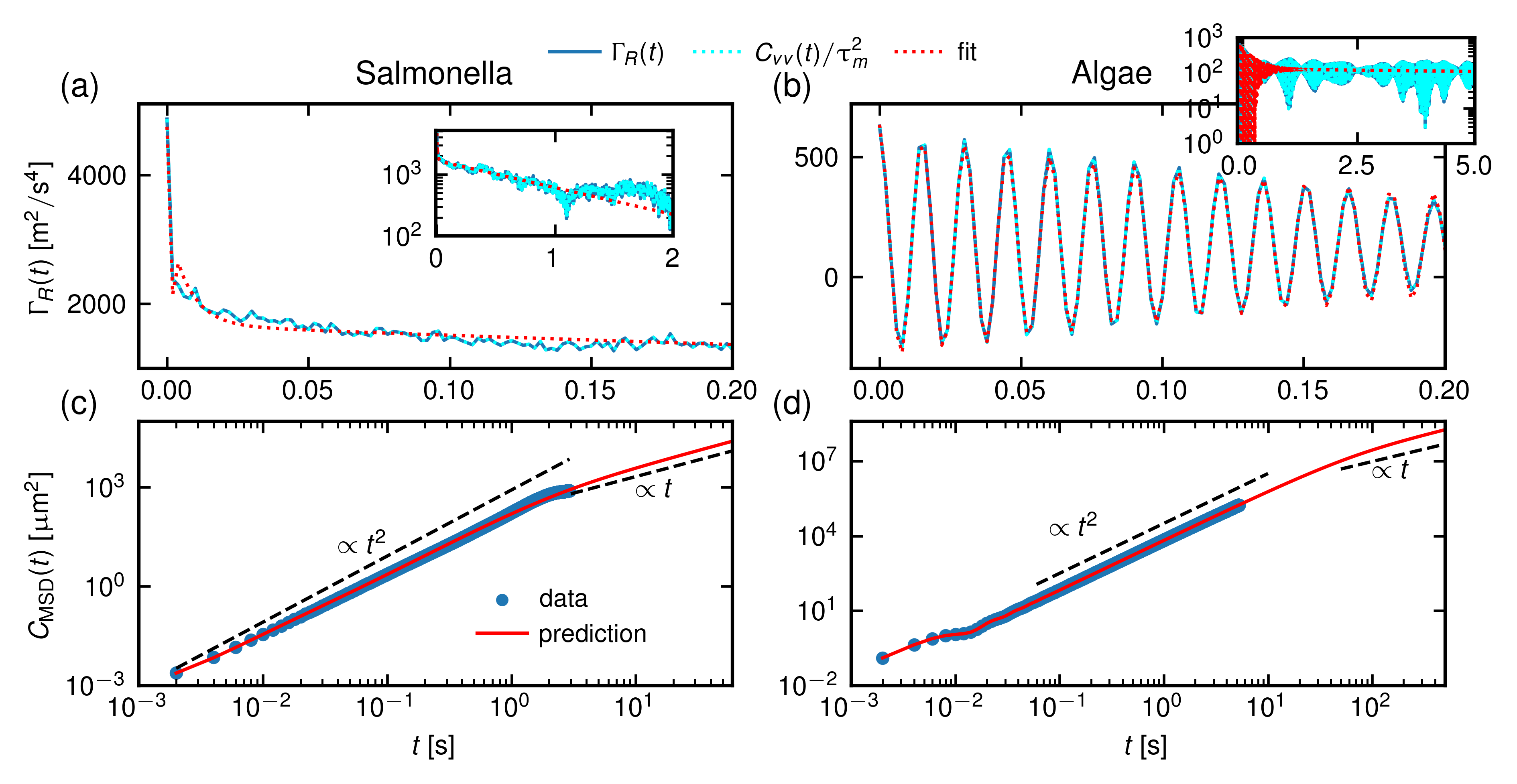}
		\caption{Propulsion-force correlation $\Gamma_R(t)$ in buffer solution with $\eta=0.89\,\rm{mPas}$ extracted via Eq.~\eqref{eq_random_force_extraction} (blue lines) is compared to $C_{vv}(t)/\tau_m^2$ (cyan dotted lines) and the respective fits of Eqs.~\eqref{eq_noneq_kern_salmo},\eqref{eq_noneq_kern_cr} (red dotted lines) are shown for (a) a single salmonella, (b) a single algal cell.
        The insets show the plots on larger time scales with logarithmic y-axis.
        Fitting details are explained in the SI.
        (c), (d) MSD of the respective single cells are shown as blue dots, where the red line is the prediction from the respective fit as explained in the SI.
        The diffusivities predicted from the fits are $D=105\,\mu\rm{m}^2/\rm{s}$ for the salmonella and $D=9.6\times 10^4\,\mu\rm{m}^2/\rm{s}$ for the algal cell.
        Black dashed lines indicate scaling behavior.
		}
		\label{fig2}
	\end{figure*}

    From individual-cell trajectories as in Fig.~\ref{fig1}c,g, we extract the propulsion-force trajectories $F_R(t)$ using Eqs.~\eqref{eq_random_force_extraction},~\eqref{eq_delta_kern}.
    The resulting force correlations for a single salmonella and a single algal cell are shown in Fig.~\ref{fig2}a,b respectively.
    The velocity autocorrelation function $C_{vv}(t)=\langle \dot{x}(0)\dot{x}(t)\rangle$ for overdamped cell motion (i.e. for $m\rightarrow 0$) is related to the force correlation by $\Gamma_R(t)=C_{vv}(t)/\tau_m^2$.
    We demonstrate that this relation holds for the salmonella and algal cells in Fig.~\ref{fig2}a,b by the cyan dotted lines ($C_{vv}(t)/\tau_m^2$ extracted independently from the data) being identical to the blue lines ($\Gamma_R(t)$ from data).
    This is expected, since the GLE Eq.~\eqref{eq_gle} for overdamped motion with the friction kernel given by Eq.~\eqref{eq_delta_kern} takes the simplified form
    \begin{equation}
    	\label{eq_le_od}
    	\dot{x}(t) = \tau_m F_R(t)\,.
    \end{equation}
    Further details on this relation are given in the SI.
    The extracted force correlations suggest for salmonella a double-exponential model
    \begin{equation}
    \label{eq_noneq_kern_salmo}
        \Gamma_R(t) = a_1 e^{-t/\tau_1} + a_2 e^{-t/\tau_2}\,,
    \end{equation}
    and for algae an oscillatory double-exponential model
    \begin{equation}
    \label{eq_noneq_kern_cr}
        \Gamma_R(t) = a_1 \cos (\Omega t) e^{-t/\tau_1} + a_2 e^{-t/\tau_2}\,.
    \end{equation}
    Fits of Eqs.~\eqref{eq_noneq_kern_salmo},~\eqref{eq_noneq_kern_cr} are shown to describe the single-cell data well in Fig.~\ref{fig2}a,b.
    These fits include effects of localization noise \cite{mitterwallner2020non,klimek_data-driven_2024,klimek_2024_cell}, which inevitably arises due to the finite pixel size of the camera and noise in the measurement, as explained in the SI.
    The propulsion-force correlations decay after a few seconds, which is much longer than the flagella oscillation period of $\sim 0.01\,\rm{s}$ for the salmonella \cite{achouri2015frequency} and $\sim 0.02\,\rm{s}$ for the algae \cite{jeanneret2016brief,mondal2021strong} and reflects a long memory and persistence in the propulsion directionality.
    This decay is slightly shorter than the thermally expected reorientation time, which indicates that the swimming machinery of the cells induces relatively small torques leading to an efficient maintenance of orientation, as explained in the SI.
    Single algal cells exhibit strong oscillations in the propulsion-force correlation at short times, which are related to the flagella oscillation period of $\sim 50\,\rm{Hz}$, as shown in Fig.\ref{fig2}b.
    In fact, residual oscillations are still present in the force correlation after a few seconds, as seen in the inset.
    These are not taken into account in the model Eq.~\eqref{eq_noneq_kern_cr} because we aim to model the mean long-time diffusivity, which is determined by the long-time exponential decay that is captured by the fit, as shown in the inset of Fig.~\ref{fig2}b.
    In contrast to other modeling approaches, we do not assume any a priori model for $\Gamma_R(t)$, rather the data determine the model to use.
    The model extracted from the data for the salmonella Eq.~\eqref{eq_noneq_kern_salmo} equals the standard active OU model in the limit of $a_1=0$.
    Therefore, the salmonella exhibit MSD behavior similar to that of an OU model, which is characterized by a ballistic regime directly transitioning to the long time diffusive regime at $\tau_2$, as shown in Fig.~\ref{fig2}c.
    The algae MSD is quite similar in the transition of ballistic to diffusive regime, but exhibits additional weak oscillations, that are visible on short times in Fig.~\ref{fig2}d.
    The data-driven models for $\Gamma_R(t)$ Eqs.~\eqref{eq_noneq_kern_salmo},~\eqref{eq_noneq_kern_cr} allow us to accurately predict the MSD by analytical relations, as explained in the SI.
    These predictions perfectly match the MSD of individual cells and extend beyond the measured temporal range, which enables us to extract the long-time diffusivity $D$ that would not be available from the direct analysis of the experimental MSD alone.
    In practice, we use the relation \cite{hansen2013theory}
    \begin{equation}
    	\label{eq_diffusion_integral}
    	D=\int_0^\infty C_{vv}(t)dt\,,
    \end{equation}
    to extract $D$ from the data using $C_{vv}(t)$ calculated with the fits in Eqs.~\eqref{eq_noneq_kern_salmo},~\eqref{eq_noneq_kern_cr}, as explained in detail in the SI.
    The mean predicted diffusivity for the algae at viscosity $\eta=0.89\,\rm{mPas}$ of $D\approx 10^5\,\mu\rm{m}^2/\rm{s}$ agrees with literature values determined by measurement of the collective cell density expanding in space over time \cite{polin2009chlamydomonas}.
    The salmonella exhibit active diffusivities on the order of $D\approx150\,\mu\rm{m}^2/\rm{s}$, which is roughly 200 times larger than the expected passive diffusion without propulsion, whereas the algae exhibit active diffusivities one million times larger than the expected passive diffusivity.
    This high increase in $D$ due to activity of the cells, paired with persistence times $\tau_2$ that are only slightly smaller than thermally driven reorientation, indicates that the cell propulsion involves a slowly relaxing cell reorientation, which is only slightly perturbed by cell propulsion,
    as explained in detail in the SI.
    \begin{figure*}
		\includegraphics{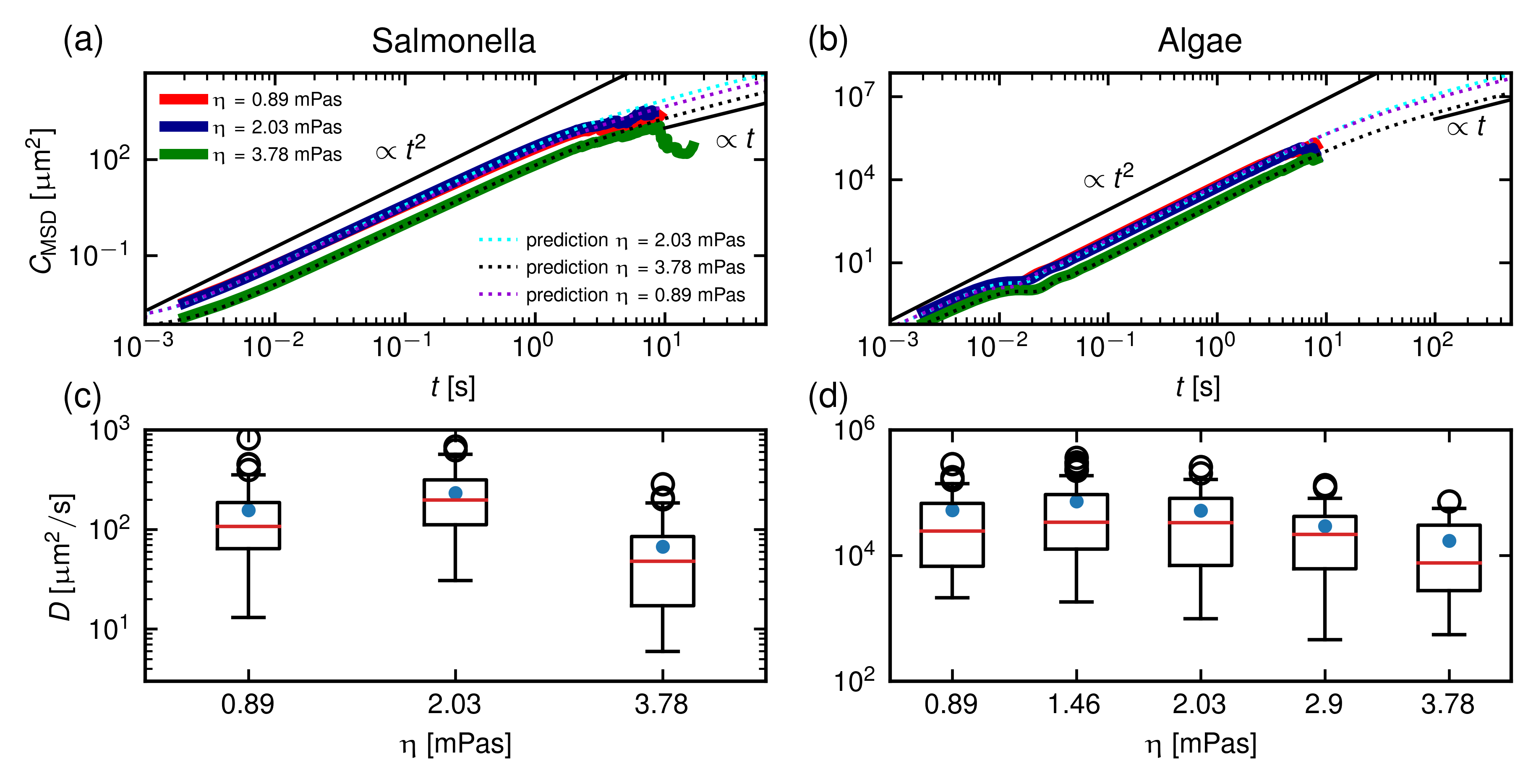}
		\caption{The ensemble average over all individual time-averaged MSDs is shown at different viscosities and the prediction by the median of all single-cell fit parameters as dotted lines for (a) salmonella, (b) algae.
        (c), (d) Boxplots of the long-time effective diffusivities that are predicted from the single-cell fits of Eqs.~\eqref{eq_noneq_kern_salmo},\eqref{eq_noneq_kern_cr} respectively, where red lines denote the median and blue dots denote the mean of the distribution.
        Boxes represent the range from the first to the third quartile and whiskers represent values within 1.5 times this interquartile range from the box edges, while black circles mark outliers beyond this range.
		}
		\label{fig3}
	\end{figure*}

    To examine the motion of the cells in different environments, we modify the solvent viscosity by systematically increasing the concentration of PEG with molecular weight $20\,\rm{kDa}$.
    We perform separate measurements at PEG concentrations ranging from $0-10\,\rm{wt}\%$ and record microscopy videos of several tens of individual cells at each concentration.
    The viscosity at each PEG concentration is determined in a separate measurement by microrheology, as explained in the SI.
    In Fig.~\ref{fig3}a,b we show the time- and ensemble-averaged MSDs and their predictions by the data-driven models at different viscosities for salmonella and algae.
    The predictions for the mean MSD result from the mean over all individual cell parameters at the given viscosity.
    Here, the functional forms of Eqs.~\eqref{eq_noneq_kern_salmo},~\eqref{eq_noneq_kern_cr} remain the same in different viscosities and only the parameters change, as shown in detail in the SI.
    The hydrodynamic friction of the cells increases linearly as a function of viscosity, as seen in Eq.~\eqref{eq_friction_hydro}.
    Thus, $\tau_m$ decreases with increasing viscosity.
    Assuming that the propulsion force is independent of the solvent viscosity would imply a monotonic decrease of velocities with increasing viscosity, as seen by Eq.~\eqref{eq_le_od}.
    Further, a viscosity-independent propulsion force could be assumed to show viscosity-independent decay times of the correlation.
    In turn, the integral over the velocity autocorrelation, which determines $D$ (seen by Eq.~\eqref{eq_diffusion_integral}), would show a monotonic decrease with increasing viscosity in this scenario.
    However, we observe a maximum in $D$ as a function of viscosity for both cell types, algae and salmonella, as seen in Fig.~\ref{fig3}c,d (and a non-monotonicity for the cell speed is shown in the SI).
    This implies that these cell types change their propulsion-force characteristics with respect to the solvent viscosity.
    Whether this change is an active adaptation by a sensory mechanism of the cells, or a consequence of the propulsion mechanism itself, requires further investigation.
    Nevertheless, this non-linear viscosity-dependence of $D$ is highly relevant for biological functions of the cells.
    For instance, salmonella are known to infect the gut and lung and cause severe symptoms from diarrhea to abdominal pain to fever \cite{dudhane2023rise,genzen2008salmonella} and these organs are protected by a mucus bio-polymer layer with high viscosity.
    Thus, the adaptation of the motility in different viscogenic environments and especially long-time diffusivity $D$ is crucial to determine how fast a salmonella can cross the mucus layer, which in turn determines the average time it needs to reach a host cell below the mucus layer.

    \section{Conclusions}
    Our data-driven modeling approach enables us to capture the short-time dynamics of cells, which is highly relevant to model cell-to-cell interactions, and at the same time to predict the long-time diffusivity, which determines how fast bacterial cells can cross large distances and potentially infect a host.
    The model perfectly matches the data for our salmonella and algae on the single-cell level and the ensemble-average level.
    
    Our asymptotic prediction of the long-time MSD behavior could be wrong if there was a long-time component present in $C_{vv}(t)$ that has no fingerprint in the observed temporal regime and sets on at later times.
    The existence of such a component is highly unlikely as endorsed by the fact that our predictions for the algae diffusivities agree with literature values from cell-suspension measurements \cite{polin2009chlamydomonas,jeanneret2016brief}.
    
    The observed maximum of the long-time diffusivity for intermediate viscosity implies that the investigated cells change their propulsion-force characteristics according to viscosity.
    Based on the observed propulsion-force correlations and mean cell speeds, our data-driven models allow us to estimate the power output of single cells, i.e. how much energy is dissipated into the environment per time.
    We find a salmonella power output in the order of $\rm{aW}$ and a algal power output in the order of $\rm{fW}$, the latter being in accordance with available literature values \cite{jones2021stochastic,ronkin1959motility}.
    Detailed tables for the calculation of the power output are given in the SI, where the power output is shown to exhibit a maximum for intermediate viscosity, similar to the diffusivity.
    This suggests that the cells actively increase the energy invested into propulsion when sensing a higher viscosity of the environment.
    We can only speculate if such an adaptation to the viscosity has evolutionary benefits or whether it is simply a byproduct of an efficient propulsion mechanism.
    The latter seems more likely because we observe the same maximum for salmonella and algae, which inhabit distinct environments and adapted to divergent evolutionary pressures.
    
    Previous studies found maxima in bacterial speed as a function of viscosity for various bacterial strains in a viscosity range around $2\,\rm{mPas}$, similar to the viscosity range where we observe maxima in $D$ for the salmonella and algae \cite{schneider1974effect}.
    However, the speed alone does not predict the diffusivity $D$, as we explain in detail in the SI.
    To determine $D$, the full parameter set of Eqs.~\eqref{eq_noneq_kern_salmo},~\eqref{eq_noneq_kern_cr} and $\tau_m$ are required.
    When determining the mean speed of a cell, it is essential to use an appropriate time interval, during which the displacement is measured.
    Usually, the instantaneous speed is approximated by the smallest available time step and is relevant to describe the short-time slope of the ballistic MSD regime.
    If the goal is to describe the long-time behavior of a cell, then the diffusivity $D$ is the more appropriate description of the dynamics, since the MSD transitions to diffusive behavior at long times and a speed does not have a clear physical meaning at times larger than the transition time $\tau_2$.

    In comparison to previous approaches \cite{mitterwallner2020non,klimek_data-driven_2024,klimek_2024_cell}, our model works for Gaussian and non-Gaussian systems.
    We note that the prediction of the friction kernel $\Gamma_v(t)$ becomes more difficult for cells that move on a substrate \cite{ron2020one,heyn2025cell}.
    Nevertheless, our modeling is applicable to any kind of cell or organism, it incorporates localization noise effects, and it is based only on the knowledge of the friction response to the environment.
    This enables us to infer intricate details about the propulsion mechanism of single cells based only on spatial trajectories of single cells in a data-driven approach, i.e. without assuming an a priori model for the propulsion-force characteristics.
    In the future, this modeling approach offers the possibility to investigate the emergence of collective effects by using the single-cell description in simulations of dense suspensions.

    \section*{Acknowledgements}
    A.K. acknowledges a useful discussion with Marco Polin at the CD25 conference.
    A.K. and R.R.N acknowledge funding by the Deutsche Forschungsgemeinschaft (DFG) through grant CRC 1449 “Dynamic Hydrogels at Biointerfaces”, Project ID 431232613, Project A03.
    P.V.B and P.S thank the ANRF grant CRG/2022/000724 and the Indian Institute of Science (IISc) for providing the support to carry out the experiments.

    \section*{Data availability}
    The data is available from the authors upon reasonable request.

	\bibliographystyle{unsrt}
	\bibliography{nonEq.bib}
    
\end{document}


\title{Supporting Information:\\Non-Markovian non-equilibrium modeling of experimental cell-motion trajectories reveals dependence of propulsion-force correlations on solvent viscosity}

    \author{Anton Klimek}
	\affiliation{Fachbereich Physik, Freie Universit{\"a}t Berlin, 14195 Berlin, Germany}
	
	\author{Prince V. Baruah}
	\affiliation{Department of Physics, Indian Institute of Science, 560012 Bangalore, India}
	
	\author{Prerna Sharma}
	\affiliation{Department of Physics, Indian Institute of Science, 560012 Bangalore, India} 
	\affiliation{Department of Bioengineering, Indian Institute of Science, 560012 Bangalore, India}
	
	\author{Roland R. Netz}
	\affiliation{Fachbereich Physik, Freie Universit{\"a}t Berlin, 14195 Berlin, Germany}
	\email{rnetz@physik.fu-berlin.de}

	\maketitle
	\tableofcontents
    \clearpage

    \section{Cell culturing and imaging of algae}
    Wild-type Chlamydomonas reinhardtii cells are inoculated from agar plates into TAP+P medium and cultured at $25\,^\circ \rm{C}$ under $12:12\,\rm{h}$ light–dark cycles on an orbital shaker operated at $137\,\rm{rpm}$.
    Cells are harvested during the logarithmic growth phase, collected in sterile $1.5\,\rm{ml}$ plastic tubes, and subsequently mixed with varying concentrations of PEG $20\,\rm{kDa}$ (Alfa Aesar).
    Observation chambers are assembled using double-sided tape spacers (Nitto Denko Corporation) with a thickness of $h=30\,\mu\rm{m}$, placed between a glass slide and a coverslip.
    To prevent non-specific adhesion of cells, both glass slides and coverslips are coated with a polyacrylamide brush.
    Imaging is performed using an inverted microscope (Olympus IX83) equipped with a high-speed CMOS camera (Phantom Miro C110, Vision Research; pixel size $5.6\,\mu\rm{m}$).
    Cells are imaged in bright-field mode using a 10x objective under red light illumination (>$610\rm{nm}$) at a frame rate of $500\,\rm{fps}$ \cite{mondal2021strong,barry2005entropy}.

    \section{Cell culturing and imaging of salmonella}
    Salmonella cells (SJW1103) are initially cultured on 2xYT agar plates ($16\,\rm{g}/\rm{l}$ tryptone, $10\,\rm{g}/\rm{l}$ yeast extract, $5\,\rm{g}/\rm{l}$ \ch{NaCl}, supplemented with 1.5\% agar) for 12 hours to grow colonies.
    $5\,\rm{ml}$ 2xYT media is inoculated with a single isolated colony and grown for $4\,\rm{h}$ at $37\,^\circ \rm{C}$ with shaking at $200\,\rm{rpm}$.
    Following growth, cells are transferred into a motility buffer consisting of $10\,\rm{mM}$ \ch{KH2PO4}, $0.1\,\rm{mM}$ EDTA, and $10 \,\rm{mM}$ sodium D-lactate, with the pH adjusted to 7.0 \cite{yamaguchi2020structural}.
    Imaging chambers are fabricated from custom-made PDMS, with a chamber height of $h=5\,\mu\rm{m}$.
    Coverslips are coated with a polyacrylamide brush to minimize non-specific cell adhesion.
    Imaging is performed on the same microscope and camera setup as for the algae, and we use a 60x phase-contrast objective under halogen lamp illumination (without a red filter) at a frame rate of $500\,\rm{fps}$.

    To produce the snapshot of a single salmonella cell with clearly visible flagella shown in the main text Fig.~1, we fluorescently label the cell body and flagella using DyLight$^{\rm{TM}}$ 550 NHS ester.
    For that, cells are washed twice in phosphate-buffered saline (PBS) by centrifugation at 3500 rcf for five minutes to remove residual growth medium.
    The dye is then added at a final concentration of $125\,\mu\rm{g}/ \rm{ml}$, and the suspension is incubated for 1 hour on a gel rocker to ensure uniform labeling.
    Following incubation, cells are again washed twice in PBS to remove unbound dye.
    Finally, fluorescence imaging is performed using an inverted microscope equipped with a 100x oil-immersion objective ($\rm{NA}\approx 1.3$) and a TRITC filter set.
    
    \section{Trajectory lengths}
\begin{figure*}
    	\includegraphics{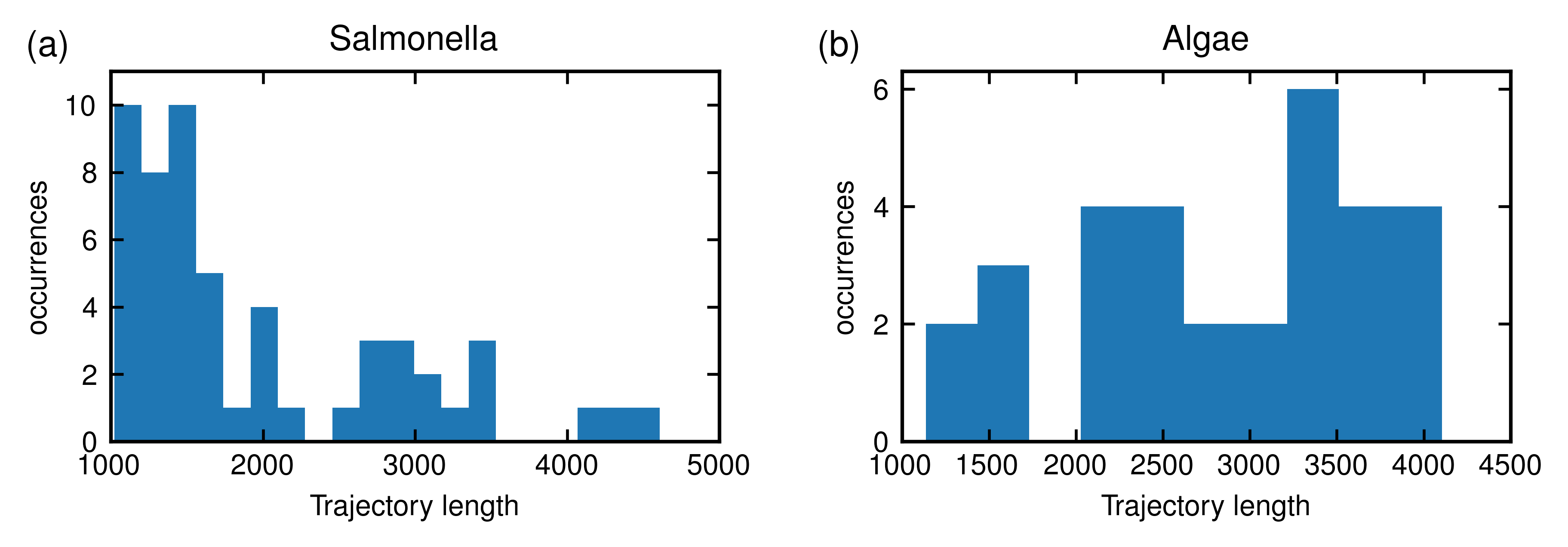}
    	\caption{Trajectory length distributions in units of time steps (with time step $\Delta=0.002\,\rm{s}$) for (a) salmonella and (b) algae.
    	}
    	\label{fig_trj_lens}
    \end{figure*}

    In Fig.~\ref{fig_trj_lens} we show the distribution of measured trajectory lengths for salmonella and algal cells in buffer with viscosity $0.89\,\rm{mPas}$.
    The length of trajectories is limited by the finite camera observation window and the finite memory storage of the camera.
    Thus, a trajectory ends either if the device memory is full or if a cell moves out of the observation window.
    A high recording frame rate leads to fast allocation of memory but is necessary to resolve the fine details of the motility patterns.
    The resulting limitation of the trajectory length leads to the lack of resolution of the long-time diffusive behavior in the MSD, as described in the man text.
    The necessity to obtain information on short as well as in long times without the ability to measure both in one experiment occurs in many different settings and is especially important for biological systems, such as cells.
    Therefore, our data-driven non-equilibrium modeling, which resolves the short-time individual-cell behavior and predicts the long-time individual-cell diffusivity, constitutes a practical approach to overcome this problem.
    
    \section{Hydrodynamic friction}
    The hydrodynamic friction coefficient of a sphere with no-slip surface boundary conditions moving at low Reynolds number is given by 
    \begin{equation}
    	\label{eq_friction_sphere}
    	\gamma_{\rm{sph}} = 3 \pi \eta d\,,
    \end{equation}
    where $\eta$ is the viscosity and $d$ the diameter of the sphere.
    For ellipsoids with rotational symmetry around the major axis $d_{\rm{maj}}$ and motion along $d_{\rm{maj}}$, Eq.~\eqref{eq_friction_sphere} is modified and takes the form \cite{kim2013microhydrodynamics,zendehroud2024molecular}
    \begin{equation}
    	\label{eq_friction_ellipsoid}
    	\gamma_{\rm{e}} = 3 \pi \eta d_{\rm{maj}} f_{\rm{e}}\,,
    \end{equation}
    where $f_{\rm{e}}$ is a modification factor given by
    \begin{equation}
    	\label{eq_factor_e}
    	f_{\rm{e}} = \frac{8 \epsilon^3}{3 (g(1+\epsilon^2) - 2\epsilon)}\,,
    \end{equation}
    with the eccentricity defined as
    \begin{equation}
    	\label{eq_excentricity}
    	\epsilon = \frac{\sqrt{d_{\rm{maj}}^2 - d_{\rm{min}}^2}}{d_{\rm{maj}}^2}
    \end{equation}
    and the logarithmic factor
    \begin{equation}
    	g=\log \left( \frac{1 + \epsilon}{1 - \epsilon} \right)\,.
    \end{equation}
    For the salmonella we find $f_{\rm{e}} \approx 0.4$ and for the algae $f_{\rm{e}} \approx 0.8$.
    The fact that $f_{\rm{e}} < 1$ means that the hydrodynamic friction of an ellipsoid with rotation symmetry around $d_{\rm{maj}}$ and moving along $d_{\rm{maj}}$ has a smaller hydrodynamic friction than a sphere of diameter $d_{\rm{maj}}$.
    Nevertheless, the friction is larger than for a sphere of diameter $d_{\rm{min}}$.
    
    A sphere that moves in the mid-plane parallel to two confining plates experiences hydrodynamic interactions with the plates, which leads to the friction \cite{ganatos1980strong}
    \begin{equation}
    	\label{eq_friction_walls}
    	\gamma_{\rm{w}} = 3 \pi \eta d f_{\rm{w}}\,,
    \end{equation}
    with
    \begin{equation}
    	\label{eq_factor_w}
    	f_{\rm{w}} = \left( 1 - 1.004 (d/h) + 0.418 (d / h)^3 + 0.21 (d / h)^4 - 0.169 (d / h)^5 \right)^{-1}\,,
    \end{equation}
    where $h$ is the distance of the confining walls.
    The hydrodynamic friction of an ellipsoid moving between two glass plates will produce slightly different interactions with the confining walls, as the flow field is different from that of a sphere.
    However, to our knowledge, there is currently no closed-form expression for the hydrodynamic friction of an ellipsoid moving between two plates.
    Therefore, we approximate the hydrodynamic interactions of our ellipsoidal-shaped cells with the walls by assuming the same increase in friction as for a sphere of diameter $d=d_{\rm{min}}$ in Eq.~\eqref{eq_factor_w}.
    This approximation becomes exact for $d_{\rm{maj}}=d_{\rm{min}}$ and is rationalized by the cells moving with their major axis oriented mainly parallel to the glass surfaces, which leads to an effective distance of the cell surface that is equivalent to that of a sphere with $d=d_{\rm{min}}$.
    Thus, combining the results of Eqs.~\eqref{eq_friction_ellipsoid},~\eqref{eq_factor_e},~\eqref{eq_factor_w} leads to the total hydrodynamic friction
    \begin{equation}
    	\label{eq_friction_hydro_SI}
    	\gamma = 3\pi \eta d_{\rm{maj}} f_{\rm{w}} f_{\rm{e}}\,,
    \end{equation}
    as stated in Eq.~(6) of the main text.
    The passive diffusivity of a cell that is not moving its flagella, for instance when the cell is dead, is given by $D_{\rm{pas}}=k_BT/\gamma$.
    The salmonella exhibit active diffusivities that are roughly 200 times larger than their passive diffusion, i.e. $D/D_{\rm{pas}}\approx 200$, whereas the algae show $D/D_{\rm{pas}}\approx 10^6$.
    These ratios are computed for the experiments in aqueous solution with viscosity $\eta=0.89\,\rm{mPas}$.
    The much higher ratio for the algae compared to the salmonella is related to the power output of the algal cells being around 1000 times larger than that of the salmonella, as follows from comparing Tabs.~\ref{tab_salmo} and \ref{tab_cr}.
    Here, a larger power output leads to a faster spatial exploration.
    
    The rotational diffusion of a passively moving sphere, which describes how fast the sphere is reorientating, is given as \cite{kim2013microhydrodynamics}
    \begin{equation}
    	\label{eq_diff_rot_sphere}
    	D_{\rm{rot}}^{\rm{sph}} = \frac{k_BT}  {\pi\eta d^3}
    \end{equation}
    Consequently, a typical reorientation time of a passively moving sphere can be estimated by
    \begin{equation}
    	\label{eq_tau_ori}
    	\tau_{\rm{ori}} = \frac{1}{D_{\rm{rot}}^{\rm{sph}} }\,.
    \end{equation}
    We use Eq.~\eqref{eq_tau_ori} to estimate typical reorientation times of cells without propulsion by using $d=d_{\rm{maj}}$ at $\eta=0.89\,\rm{mPas}$.
    This yields $\tau_{\rm{ori}}\approx15\,\rm{s}$ for spheres with comparable size to the salmonella and $\tau_{\rm{ori}}\approx680\,\rm{s}$ for the algae.
    These reorientation times are roughly 10 and 30 times higher than the long-time decay of the propulsion force correlation $\tau_2$ for the respective cells.
    This suggests that the active motion promotes faster reorientation compared to a thermally driven reorientation, presumably due to angular fluctuation of the flagella with respect to the cell body \cite{jeanneret2016brief}.
    
    From a different viewpoint, the increased translational diffusion of the actively moving cells compared to passive diffusion originates from the propulsion forces being much larger than the forces due to thermal fluctuations.
    This is sometimes interpreted by an effective temperature \cite{cugliandolo2011effective}, which would then be $D/D_{\rm{pas}}$ times higher than the temperature in the experiment.
    In this picture of an effectively increased temperature due to activity of the cell, one would expect from Eq.~\eqref{eq_tau_ori} that the active orientation time $\tau_{\rm{ori}}^{\rm{act}}=\tau_{\rm{ori}} \frac{D_{\rm{pas}}}{D}$ is 200 times smaller than the passive one for salmonella and one million times smaller for the algae; i.e. one would expect $\tau_{\rm{ori}}^{\rm{act}}\approx75\,\rm{ms}$ for the salmonella and $\tau_{\rm{ori}}^{\rm{act}}=0.68\,\rm{ms}$ for the algae.
    The persistence time of the cells $\tau_2$ signals the transition from ballistic motion due to reorientations and is thus related to the reorientation time.
    $\tau_2$ is much longer than $\tau_{\rm{ori}}^{\rm{act}}$, where $\tau_2$ values are shown in Fig.~\ref{fig_param_distr}b,f.
    This implies that the cells increase their long-time diffusivity by their active propulsion while exerting relatively small torques that lead to only a small decrease of orientational persistence compared to thermally induced change of orientation.
    Especially the algal cells show very long persistence times paired with high diffusivity, which suggests that their propulsion forces induce mainly forward motion and little orientational drift.
    
    $D_{\rm{pas}}$ is also used in the micro-rheology measurements to determine the viscosity $\eta$ from the diffusion of polystyrene spheres of diameter $d=2.19\,\mu\rm{m}$ in the solvents with different PEG concentrations.
    Here, the MSD is determined from single-particle tracking of identical spheres, which allows us to extract $\eta$ at the experimental temperature $T=298\,\rm{K}$ via Eq.~\eqref{eq_friction_sphere}.

    \section{Fitting details}
    \label{sec_fitting}
    Fits of the propulsion force correlation $\Gamma_R(t)$ by the model suggested by the data Eqs.~(8),~(9) in the main text are performed by taking into account the finite temporal resolution and localization noise as described in \cite{mitterwallner2020non,klimek_2024_cell,klimek_data-driven_2024}.
    Any experimentally observed position of a cell $x^{\rm{exp}}(t)$ includes an uncertainty due to a finite spatial resolution and noise in the experimental setup.
    This uncertainty can be modeled by uncorrelated Gaussian noise on the position
    \begin{equation}
    	\label{eq_pos_noise}
    	x^{\rm{exp}}(t) = x(t) + \xi (t)\,,
    \end{equation}
    where $\sigma_{\rm{loc}}$ defines the localization noise width $\langle \xi (0)\xi (t) \rangle = \delta_{0,t}\sigma_{\rm{loc}}$ and $\delta_{0,t}$ is 1 if $t=0$ and zero elsewise.
    Eq.~\eqref{eq_pos_noise} leads to the noisy MSD
    \begin{equation}
    	\label{eq_msd_noise}
    	C_{\textrm{MSD}}^{\rm{noise}}(t)=C_{\textrm{MSD}}^{\rm{theo}}(t)+2(1-\delta_{t,0})\sigma_{\textrm{loc}}^2 \,.
    \end{equation}
    In order to perform a model fit, we need to find a model for the theoretical MSD in the absence of noise $C_{\textrm{MSD}}^{\rm{theo}}(t)$.
    For this, we use the relations
    \begin{equation}
    	\label{eq_force_vacf}
    	\Gamma_R(t)=C_{vv}(t)/\tau_m^2
    \end{equation}
    and
    \begin{equation}
    	\label{eq_msd_vacf_relation}
    	C_{vv}(t)=\frac{1}{2}\dfrac{\rm{d}^2}{\rm{d}t^2}C_{\rm{MSD}}(t)\,,
    \end{equation}
    where Eq.~\eqref{eq_force_vacf} is a consequence of the cell motion being overdamped on the experimentally resolved time scales, which is reflected by inertial times $\tau_m$ that are much smaller than the experimental time step $\Delta$, and Eq.~\eqref{eq_msd_vacf_relation} holds for every stationary process.
    Eq.~\eqref{eq_msd_vacf_relation} implies
    \begin{equation}
    	\label{eq_vacf_msd_integral}
    	C_{\rm{MSD}}(t)= 2 \int_0^t dt'\int_0^{t'}dt'' C_{vv}(t'')\,.
    \end{equation}
    Thus, the models suggested by the data for $\Gamma_R(t)$ in Eqs.~(8),~(9) directly translate into models for the MSD by using Eqs.~\eqref{eq_force_vacf} and \eqref{eq_vacf_msd_integral}.
    The discretized version of Eq.~\eqref{eq_msd_vacf_relation} reads
    \begin{equation}
    	\label{eq_vacf_fit}
    	C_{vv}^{\rm{fit}}(i \Delta) = \frac{C_{\textrm{MSD}}^{\rm{noise}}((i+1)\Delta)-2C_{\textrm{MSD}}^{\rm{noise}}(i \Delta)+C_{\textrm{MSD}}^{\rm{noise}}((i-1)\Delta)}{2\Delta^2} \,
    \end{equation}
    and is used to fit the data by a least square fit minimizing the cost function
    \begin{equation}
    	\label{eq_cost_function}
    	E_{\textrm{cost}}=\sum_{i=0}^{n-1} (C_{vv}^{\rm{exp}}(i \Delta)-C_{vv}^{\rm{fit}}(i \Delta))^2\,,
    \end{equation}
    as explained in detail in \cite{mitterwallner2020non,klimek_2024_cell,klimek_data-driven_2024}.
    We find localization noise widths in the order of $\sigma_{\rm{loc}}\sim0.01\,\mu\rm{m}$ for the salmonella and $\sigma_{\rm{loc}}\sim0.05\,\mu\rm{m}$ for the algae, which is roughly 20\% of a typical displacement per time step $\Delta$.
    A high localization noise yields overestimated displacements per time step as seen by Eq.~\eqref{eq_msd_noise}, which would lead to an overestimation of the diffusivity unless accounted for by our fitting procedure in Eq.~\eqref{eq_cost_function}.
    
    The GLE Eq.~(2) in the main text for the instantaneous friction $\Gamma_v(t)=2\delta(t)/\tau_m$ leads to the equation of motion
    \begin{equation}
    	\label{eq_le}
    	\ddot{x}(t)=-\dot{x}(t)/\tau_m + F_R(t)\,,
    \end{equation}
    which, in the observed overdamped limit with small $\tau_m$, becomes
    \begin{equation}
    	\label{eq_le_od}
    	\dot{x}(t) = \tau_m F_R(t)\,.
    \end{equation}
    From Eq.~\eqref{eq_le_od} it is directly seen that $\langle v(t)v(t')\rangle = \tau_m^2 \langle F_R(t)F_R(t')\rangle$.
    
\begin{figure*}
    	\includegraphics{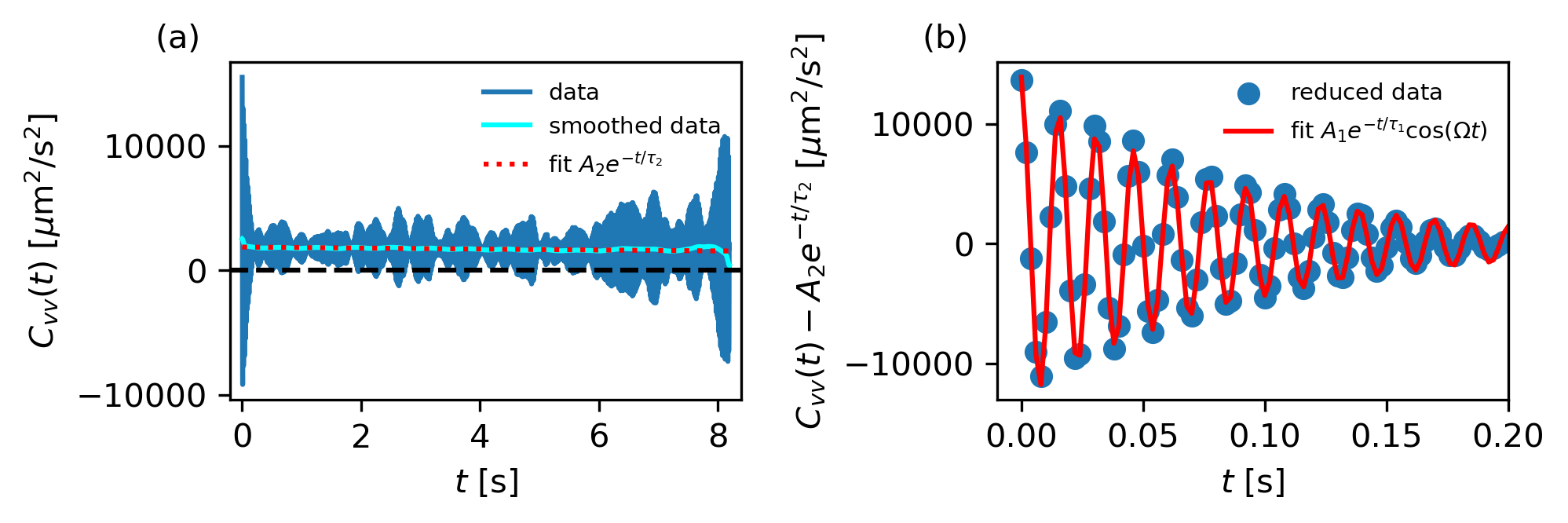}
    	\caption{(a) The VACF of a single algal cell (blue line) is smoothed with a Gaussian of width 20 time steps (cyan line).
    		The red dotted line represents the least-squares fit of the long-time exponential contribution to the smoothed data, where $A_2=a_2\tau_m^2$ (see Eq.~(9) of the main text) and the localization noise is set to zero.
    		The long decay time is fitted as $\tau_2\approx 45\,\rm{s}$ and the dashed horizontal line shows the zero.
    		(b) The fit shown in (a) is subtracted from the $C_{vv}(t)$ data shown in (a), which results in the reduced data shown as blue dots.
    		This reduced data set is then fitted by the short-time decaying oscillation including localization noise, which leads to the red line.
    		Here, $A_1=a_1\tau_m^2$ (see Eq.~(9) of the main text).
    	}
    	\label{fig_fit_explain}
    \end{figure*}

    To fit the model of Eq.~(9) to the single algal cell data, we perform two separate fits; one of the short time behavior and one of the long time decay.
    First, we smooth the VACF by a Gaussian of width 20 time steps, which means a width of $0.04\,\rm{s}$.
    The smoothed data contains no oscillations and reflects the long-time decay, as seen in Fig.~\ref{fig_fit_explain}a.
    We fit the long-time contribution of the model $A_2e^{-t/\tau_2}$ to the smoothed data in the time interval between $0-3\,\rm{s}$ using the fit described above, where we set the localization noise width $\sigma_{\rm{loc}}$ to zero.
    Subsequently, we subtract the fitted long-time contribution from the data and fit the short time contribution $A_1e^{-t/\tau_1}\cos (\Omega t)$ to the reduced data, as shown in Fig.~\ref{fig_fit_explain}b.
    Here, we include the localization noise in the fit to the short time contribution, which we fit in the time interval between $0-0.2\,\rm{s}$.
    Finally, the two fitted components are added to yield $C_{vv}(t)= A_1e^{-t/\tau_1}\cos (\Omega t) + A_2e^{-t/\tau_2}$.
    This final fit result is shown for a single algal cell in Fig.~2b in the main text.
    
	\section{Parameter distributions}
\begin{figure*}
		\includegraphics{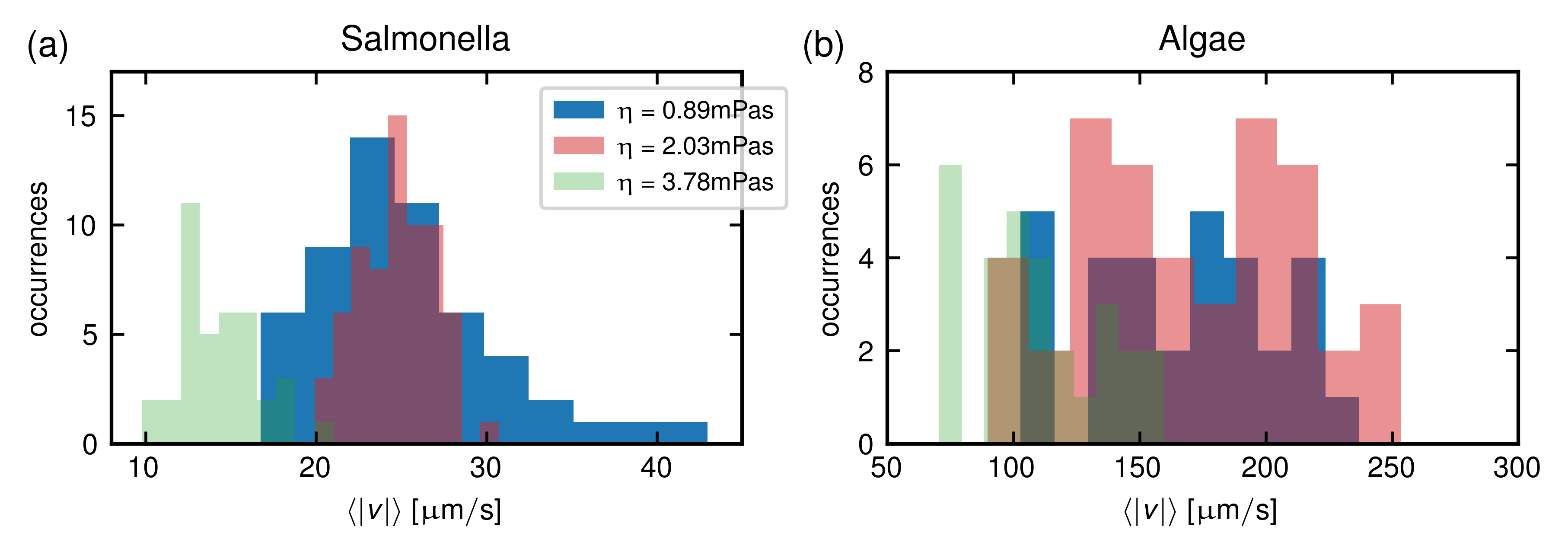}
		\caption{Distribution of mean speed $|v|=\sqrt{v_x^2 + v_y^2}$ directly extracted from the data and shown for different viscosities for (a) salmonella, (b) algae.
		}
		\label{fig_distr_speed}
	\end{figure*}

\begin{figure*}
		\includegraphics{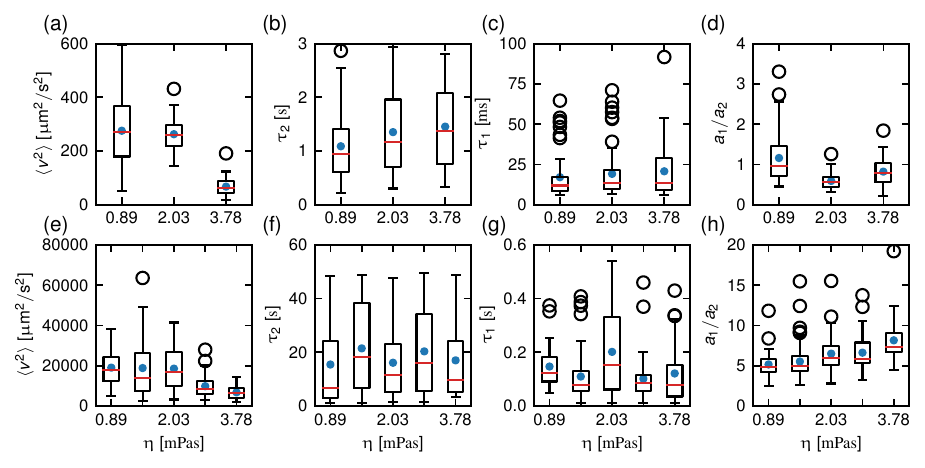}
		\caption{Boxplots for mean squared velocity $\langle v^2 \rangle$, longest relaxation time $\tau_2$, shortest resolved relaxation time $\tau_1$ and ratio of amplitudes from short-time contribution and long-time contribution to the propulsion force correlation $a_1/a_2$ for (a)-(d) salmonella and (e)-(h) algae.
		}
		\label{fig_param_distr}
	\end{figure*}

	In Fig.~\ref{fig_distr_speed} we show the mean speed distributions for single cells for three different viscosities.
	Velocities at each time point of the trajectory are computed as
	\begin{equation}
		\label{eq_velocity_finite}
		v_{i+1/2} = \frac{x_{i+1} - x_i}{\Delta}\,,
	\end{equation}
	where $x_i$ is the position at time $i\Delta$ and $\Delta$ is the recording time-step.
	The speed distributions shown in Fig.~\ref{fig_distr_speed} are created by averaging over all values of an individual trajectory and then creating a histogram of these mean individual trajectory speeds.
	For the salmonella, the distributions shift to smaller speeds with increasing viscosity, whereas this shift is only clearly visible for the highest viscosity in the case of the algae.
	Figs.~\ref{fig_param_distr}a,e show that the mean squared velocity mainly decreases with increasing viscosity beyond the value of $\eta = 2.03\,\rm{mPas}$. For lower viscosities, the mean squared velocity is rather constant.
	In contrast, the longest decay time $\tau_2$ increases with increasing viscosity for the salmonella and shows no clear trend for the algae, as seen in Figs.~\ref{fig_param_distr}b,f.
	
	The long time diffusion constant is related to $C_{vv}(t)$ via \cite{hansen2013theory}
	\begin{equation}
		\label{eq_diff_vacf}
		D=\int_0^\infty C_{vv}(t) dt\,.
	\end{equation}
	This relation can be understood when inserting $C_{vv}(t)=\frac{1}{2}\dfrac{\rm{d}^2}{\rm{d}t^2}C_{\rm{MSD}}(t)$, which holds for stationary observables, into Eq.~\eqref{eq_diff_vacf}.
	This yields $D=\frac{1}{2}\dfrac{\rm{d}}{\rm{d}t}C_{\rm{MSD}}(t)|_0^\infty$.
	Due to the symmetry of the MSD, it follows that $\dfrac{\rm{d}}{\rm{d}t}C_{\rm{MSD}}(0) = 0$ and the slope of the MSD in the infinite-time limit is given by $\dfrac{\rm{d}}{\rm{d}t}C_{\rm{MSD}}(\infty) = 2D$, which defines the long-time diffusivity $D$.
	We predict $D$ from the data by
	\begin{equation}
		\label{eq_diff_salmo}
		D_{\rm{sal}} = A_1 \tau_1 + A_2 \tau_2
	\end{equation}
	for the salmonella and
	\begin{equation}
		\label{eq_diff_cr}
		D_{\rm{alg}} = \frac{A_1 \tau_1}{1 + \tau_1^2\Omega^2} + A_2 \tau_2
	\end{equation}
	for the algae.
	Eqs.~\eqref{eq_diff_salmo} and \eqref{eq_diff_cr} result from the analytical integration (Eq.~\eqref{eq_diff_vacf}) of the fitting functions Eqs.~(8),~(9) respectively and using Eq.~\eqref{eq_force_vacf}, where we define $A_i=a_i\tau_m^2$.
	
	In both Eqs.~\eqref{eq_diff_salmo} and \eqref{eq_diff_cr}, the long-time contribution to the diffusivity $A_2\tau_2$ is much larger (roughly by a factor of hundred) than the respective short time contribution.
	This is understood by Figs.~\ref{fig_param_distr}b-d,~f-h, where $\tau_2$ is hundred time larger than $\tau_1$, $A_1/A_2$ is in the order of one and $\tau_1\Omega >> 1$ (as follows from the fact that many oscillations are seen in Fig.~\ref{fig_fit_explain}a).
	The non-monotonicity of $D$ with increasing viscosity, that we show in the main text Fig.~3, can thus be interpreted as the result of decreasing velocities competing with increasing persistence times, since the integral in Eq.~\eqref{eq_diff_vacf} is dominated by the term $A_2\tau_2$.
	In other words, the cells exhibit longer ballistic regimes in the MSD but move at lower speeds.
	If the propulsion forces were the same in different viscosities, one would expect $D$ to decrease monotonically as a function of viscosity, as explained in section~\ref{sec_functional_form}.
	The fact that there is a maximum in the diffusivity reveals that the cells adjust their propulsion-force characteristics to the surrounding viscosity.
	Further research is needed to determine whether this adjustment of propulsion is simply a side effect of changing the viscosity or if it has a biological purpose.
	Finally, the values of $\tau_2$ at $\eta=0.89\,\rm{mPas}$ in Fig.~\ref{fig_param_distr}f for the algal cells are comparable to reorientation times reported in the order of $\sim 10\,\rm{s}$ \cite{liu2020swimming}.

    \section{Functional form of propulsion force correlation is independent of viscosity}
    \label{sec_functional_form}
\begin{figure*}
		\includegraphics{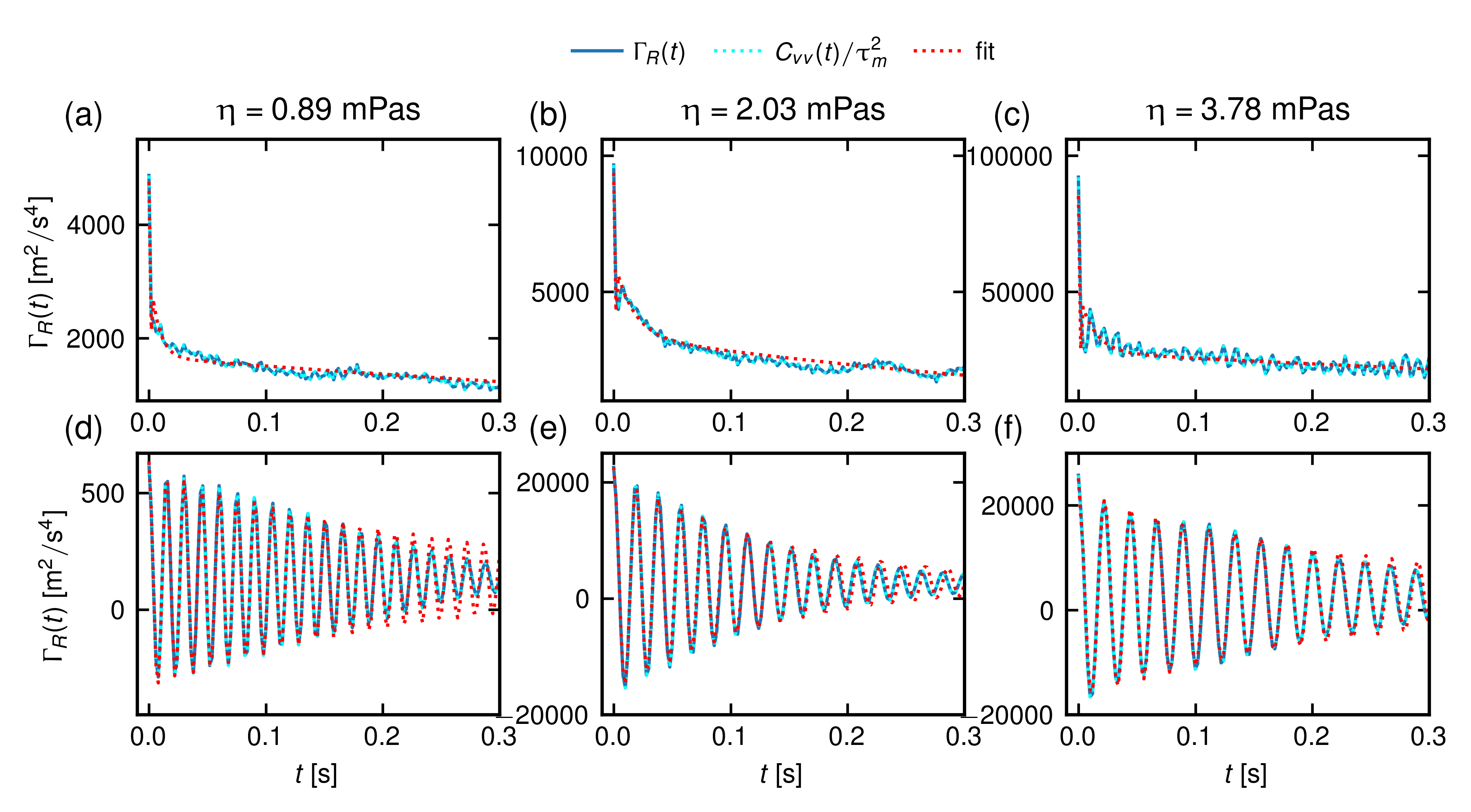}
		\caption{Propulsion force correlation $\Gamma_R(t)$ (a)-(c) for individual salmonella and (d)-(f) for individual algal cells at different viscosity given in the titles.
		}
		\label{fig_functional_form}
	\end{figure*}

    The functional forms of $\Gamma_R(t)$ that the salmonella and algae exhibit, Eqs.~(8),~(9) respectively, do not change with increasing viscosity, as shown in Fig.~\ref{fig_functional_form}.
    However, the parameters $a_i$, $\tau_i$ and $\Omega$ change as a function of viscosity, as shown in Fig.~\ref{fig_param_distr}.
    The Markovian friction $\Gamma_v(t)=2\delta(t)/\tau_m$ in its discrete form $\Gamma_v^0=2/(\tau_m\Delta)$ and $\Gamma_v^i=0$ for $i>0$ leads to the expression for the friction force from Eq.~(4) in the main text
    \begin{equation}
        \label{eq_force_stokes}
        F_R^i = \ddot{x}^i + \dot{x}^i/\tau_m\,.
    \end{equation}
    Since the cell motion is overdamped, which means that $\tau_m$ is very small, Eq.~\eqref{eq_force_stokes} simplifies to
    \begin{equation}
        \label{eq_force_extract_od_stokes}
        F_R^i = \dot{x}^i/\tau_m\,.
    \end{equation}
    For all our data, we use Eq.~\eqref{eq_force_stokes} to extract the propulsion force trajectories and then compute the autocorrelation by Fourier methods to determine $\Gamma_R(t)$.
    However, we show in the main text and in Fig.~\ref{fig_functional_form} that $\Gamma_R(t)=C_{vv}(t)/\tau_m$, which proves that inertial terms are in fact negligible and Eq.~\eqref{eq_force_extract_od_stokes} would also be appropriate to extract the forces.

    The hydrodynamic friction $\gamma$ increases linearly with the viscosity, as seen in Eq.~\ref{eq_friction_hydro_SI}.
    Thus, $\tau_m=m/\gamma$ decreases monotonically as a function of viscosity.
    If the propulsion forces of the cells were independent of the solvent viscosity, one would expect from Eq.~\eqref{eq_force_extract_od_stokes} to observe a monotonic decrease of velocities as a function of $\eta$.
    We observe that this is not the case and mean squared velocities show a slight non-monotonicity for the investigated salmonella and algae, as seen in Fig.~\ref{fig_param_distr}a,e and Fig.~\ref{fig_distr_speed}.
    Moreover, a monotonic decrease of velocity would imply a monotonic decrease of long-time diffusivity $D$ if the decay time of $C_{vv}(t)$ was independent of $\eta$, since $D$ is related to the integral over $C_{vv}(t)$, seen in Eq.~\eqref{eq_diff_vacf}.
    Again, we observe a non-monotonicity marked by the maximum of $D$ as a function of viscosity $\eta$ shown in the main text Fig.~3c,d, which further contradicts the assumption of a viscosity-independent propulsion force.
    Thus, we conclude that the propulsion force is dependent on the solvent viscosity, which is in line with the finding of a maximum of the single-cell power output at the same viscosity, shown in Tabs.~\ref{tab_salmo},~\ref{tab_cr}.

    \section{Propulsion force is Gaussian and stationary}
    \label{sec_gaussian_force}
\begin{figure*}
		\includegraphics{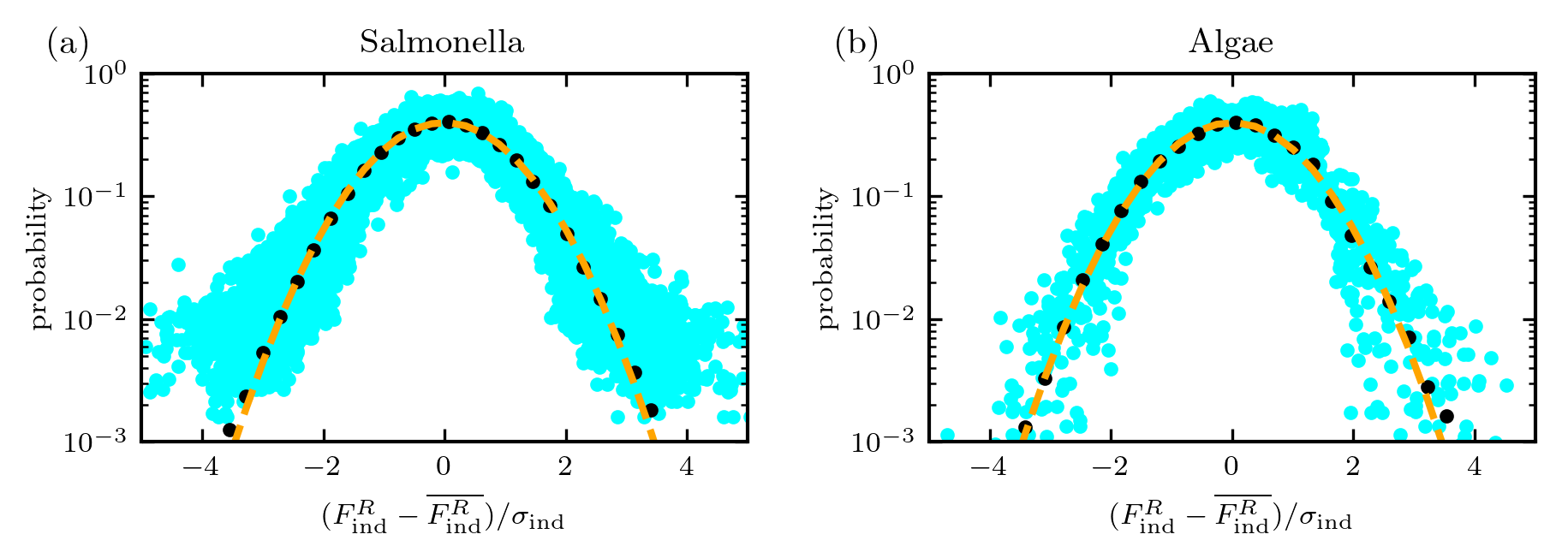}
		\caption{Distribution of propulsion forces rescaled by subtracting the mean over the individual trajectory and dividing by the variance of the individual trajectory. Distributions for (a) 55 single salmonella cells and (b) 31 single algal cells are shown as cyan dots, the black dots represent the mean over all cyan dots and the orange dashed line shows the standard normal distribution.
		}
		\label{fig_gauss_explain}
	\end{figure*}

    We extract propulsion forces from trajectories using Eq.~(4) in the main text, which takes the form of Eq.~\eqref{eq_force_stokes} for $\Gamma_v(t)=2\delta(t)/\tau_m$, as explained in section~\ref{sec_functional_form}.
    The extracted forces $F_R^i$ are distributed according to a Gaussian distribution for every single cell.
    This is shown in Fig.~\ref{fig_gauss_explain}, where we subtract the individual trajectory mean $\bar{F}_{\rm{ind}}^R$ and divide by the individual trajectory standard deviation $\sigma_{\rm{ind}}$, which results in the standard normal distribution of mean zero and width one for every single cell.

\begin{figure*}
		\includegraphics{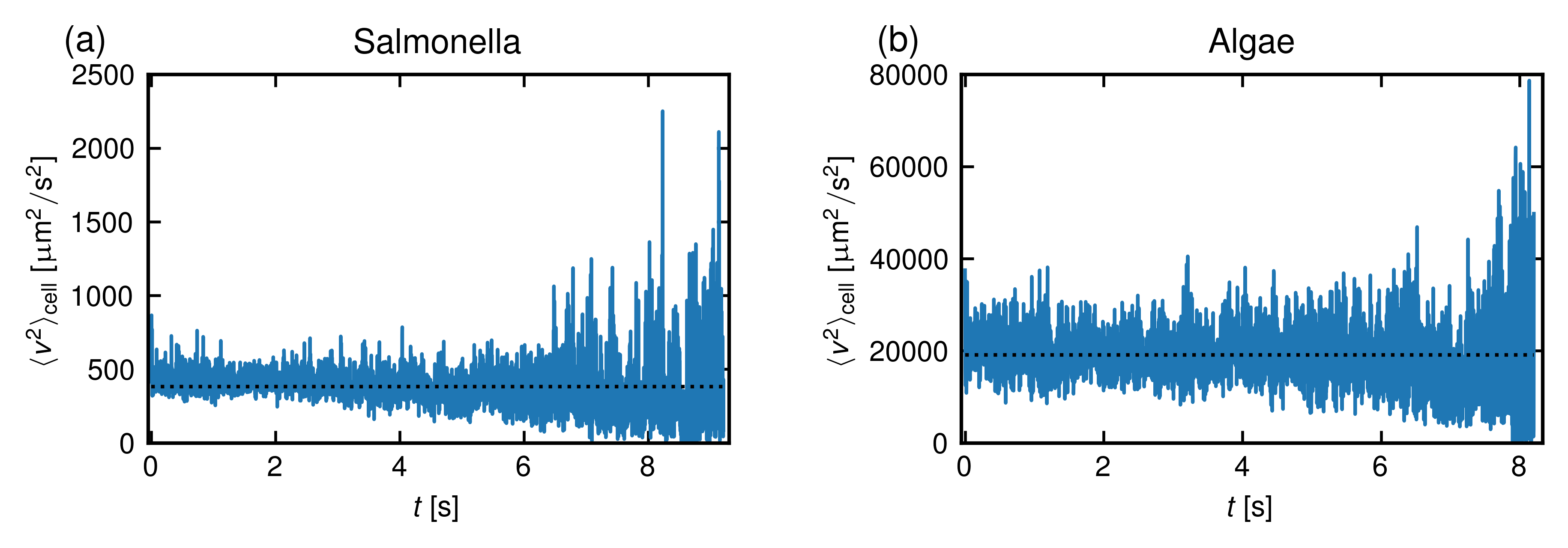}
		\caption{The ensemble-averaged squared velocity (mean over all cells) is shown as a function of time for (a) salmonella and (b) algae.
		The black dotted lines indicate the constant value of the time average over the ensemble-averaged squared velocity.
		}
		\label{fig_stationarity}
	\end{figure*}

	 Fig.~\ref{fig_stationarity} shows the ensemble averaged squared velocity at $\eta=0.89\,\rm{mPas}$ as a function of time, where velocities are extracted by Eq.~\eqref{eq_velocity_finite}.
	 Fluctuations in this mean squared velocity are due to an average over a finite number of cells with individual differences.
	 As demonstrated in Fig.~\ref{fig_trj_lens}, only few cells exhibit trajectories longer than $6\,\rm{s}$.
	 Therefore, the ensemble average at such long times becomes increasingly noisy.
	 Nevertheless, the black dotted lines indicate that the ensemble average remains constant within the experimentally observed times.
	 Further, Eq.~\eqref{eq_le_od} shows that the propulsion force is proportional to the velocity, which implies that the ensemble-averaged squared propulsion force is constant in time as well.
	 This means that the cell motion is a stationary process and Eq.~\eqref{eq_msd_vacf_relation} is valid.
	
    \section{Power output of single cells}
    The force correlation at time zero according to the data-driven fitting functions Eqs.~(8),~(9) is given by
    \begin{equation}
    \label{eq_force_corr_zero}
        \Gamma_R(0) = a_1 + a_2\,.
    \end{equation}
    In Fig.~\ref{fig_gauss_explain} we show that the propulsion force distribution is Gaussian for every single cell.
    For Gaussian one-dimensional observables the expectation value of the absolute value is related to the variance as $\langle|F_R(t)|\rangle = \sqrt{\frac{2}{\pi} \langle F_R^2(t) \rangle}$, which leads to
    \begin{equation}
    \label{eq_gauss_absolute_2d}
        \langle|F_R(t)|\rangle = \sqrt{\frac{4}{\pi} \langle F_R^2(t) \rangle}
    \end{equation}
    in our case of two-dimensional observables.
    The propulsion force values given in tables~\ref{tab_salmo},~\ref{tab_cr} are extracted from the data by combining Eqs.~\eqref{eq_force_corr_zero} and \eqref{eq_gauss_absolute_2d}.
    Similarly, the expectation value of the speed is extracted using $\langle v^2\rangle = A_1 + A_2 = (a_1 + a_2)/\tau_m^2$ and the equivalent of Eq.~\eqref{eq_gauss_absolute_2d} for the velocity.
    Finally, the power output is estimated by $P=\langle m |F_R(t)|\rangle \langle |v| \rangle$, which is equivalent to the assumption that the friction force balances the net-forward propulsion force.
    Here, the cell mass is estimated by $m=\rho V$ with $\rho=1000\,\rm{kg}/m^3$ the density of water and $V=\frac{\pi}{6}d_{\rm{min}}^2 d_{\rm{maj}}$ is the volume of the ellipsoidal-shaped cells with major axis $d_{\rm{maj}}$ and minor axis $d_{\rm{min}}$.
    The average mass of the salmonella is $m\sim 10^{-15}\,\rm{kg}$ and for the algae $m\sim 3 \times 10^{-13}\,\rm{kg}$; the latter is comparable to the directly measured cell weight of $m\sim10^{-14}\,\rm{kg}$ \cite{chioccioli2014flow}.
    These values of $m$ are independent of the solvent viscosity.
    
    Using the fitted parameters $a_1$ and $a_2$ to estimate the power output has the advantage that the localization noise effects are included.
    In contrast, directly extracting the values for $\Gamma_R(0)$ and $\langle |v| \rangle$ from the data would overestimate the power output, as localization noise leads to increased estimates for both, the instantaneous speed and the force amplitude \cite{klimek_2024_cell,klimek_data-driven_2024}.
    It is important for an accurate estimate of the power output to use the value for the instantaneous speed, by which we mean the speed at an infinitely small time interval.
    Experimentally this means to use a small recording time-step that compromises between a good ratio of localization noise to actually moved distance and short enough time step to resolve the speed at the smallest moved spacial increments.
    For this, we use a recording rate of $500\,\rm{fps}$.
    Measuring the speed of micro-organisms at longer temporal scales, for instance by the traveled distance over the span of one second, leads to an underestimation of the speed of the organism on shorter time and therefore, to an underestimation of the power output of the organism.
    This effect becomes especially prominent for recording time-steps that are larger than the transition time from the short-time ballistic behavior to the long-time diffusive behavior in the MSD.

    \begin{table}
    \centering
    \begin{tabular}{|c|c|c|c|}
    \hline
    $\eta$ $[\rm{mPa}\,\rm{s}]$ & $\langle m |F_R(t)|\rangle [\rm{fN}]$ & $\langle |v| \rangle  [\frac{\mu\rm{m}}{\rm{s}}]$ & $P$ $[aW]$  \\
    \hline
    0.89 & 119 & 19 & 2.2 \\
    \hline
    2.03 & 197 & 18 & 3.6 \\
    \hline
    3.78 & 195 & 9 & 1.8 \\
    \hline
    \end{tabular}
    \caption{Mean force amplitude, speed and estimated power output in different viscosities averaged over all salmonella.
    The multiplication of $F_R(t)$ by the cell mass $m$ yields units of a force, where $m\sim  10^{-15}\,\rm{kg}$ is estimated by the mean volume of all individual cells using the density of water $\rho=1000\,\rm{kg}/\rm{m}^3$.
	Here $m$ is independent of $\eta$.}
    \label{tab_salmo}
    \end{table}

    \begin{table}
    \centering
    \begin{tabular}{|c|c|c|c|}
    \hline
    $\eta$ $[\rm{mPa}\,\rm{s}]$ & $\langle m |F_R(t)|\rangle$ $[\rm{pN}]$ & $\langle |v| \rangle$ $[\frac{\mu\rm{m}}{\rm{s}}]$ & $P$ $[fW]$  \\
    \hline
    0.89 & 12 & 156 & 1.9 \\
    \hline
    1.46 & 23 & 155 & 3.6 \\
    \hline
    2.03 & 30 & 154 & 4.7 \\
    \hline
    2.90 & 32 & 112 & 3.6 \\
    \hline
    3.78 & 37 & 93 & 3.5 \\
    \hline
    \end{tabular}
    \caption{Mean force amplitude, speed and estimated power output for different viscosities averaged over all algae.
    The force amplitude and speed are estimated by the fit parameters, such that the effects of localization noise are properly taken into account.
    The multiplication of $F_R(t)$ by the cell mass $m$ yields units of a force, where $m\sim 3 \times 10^{-13}\,\rm{kg}$ is estimated by the mean volume of all individual cells using the density of water $\rho=1000\,\rm{kg}/\rm{m}^3$.
	Here $m$ is independent of $\eta$.}
    \label{tab_cr}
    \end{table}

	\section{Force correlation for oscillating friction kernel}
\begin{figure*}
		\includegraphics{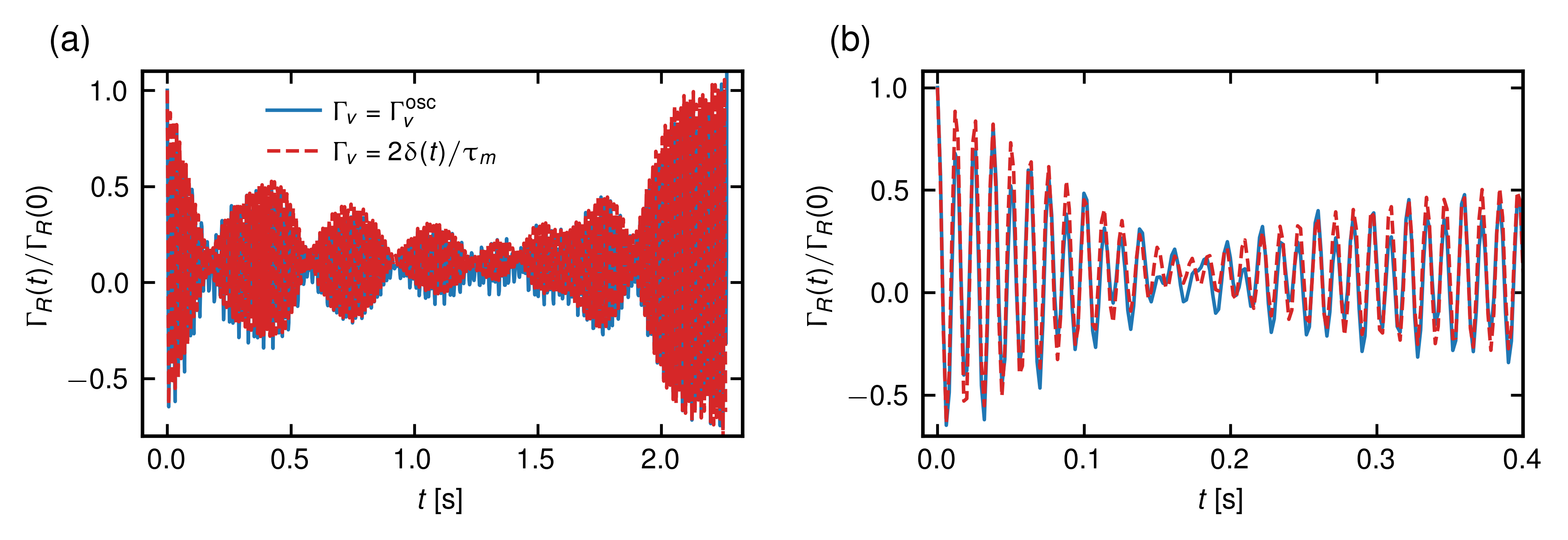}
		\caption{The propulsion force correlation $\Gamma_R(t)$ of a single algal cell extracted under the assumption of frequency-independent Stokes friction $\Gamma_v(t)=2\delta(t)/\tau_m$ (Eq.~(5) main text) is shown as red dashed line.
			In comparison the blue line is the extracted force correlation using $\Gamma_v(t)=\Gamma_v^{\rm{osc}}(t)$ in Eq.~\eqref{eq_gammav_osc} with the frequency chosen as the flagella beat frequency of $\Omega=2\pi\, 50\,\rm{Hz}$.
			(a) Full length of the data, (b) short time behavior up to $0.2\,\rm{s}$.
		}
		\label{fig_gammar_diff_gammav}
	\end{figure*}

	The assumption of frequency-independent hydrodynamic friction is based on the stationary hydrodynamic description of rigid spherical objects in water \cite{kiefer2025effect,kim2013microhydrodynamics}, but does not include effects due to the time dependent shape of the flagella.
	For the salmonella, the flagella are dragged along behind the cell body.
	In contrast, the flagella of the algae extend sidewards during the breast-stroke motion, which will periodically increase the hydrodynamic friction of the cell during that time.
	Thereby, not only the total friction but also the time dependence of $\Gamma_v(t)$ could be relevant.
	We test the influence of an additional oscillating friction
	\begin{equation}
		\label{eq_gammav_osc}
		\Gamma^{\rm{osc}}_v(t)=\frac{\Omega}{2\tau_m}\cos (\Omega t) + 2\delta(t)/\tau_m\,,
	\end{equation}
	motivated by the oscillatory motion of the flagella, which potentially increases and decreases the hydrodynamic friction periodically.
	The oscillatory prefactor component of the form $\frac{\Omega}{2\tau_m}$ is chosen such that the total friction $G(t)=\int_0^t \Gamma_v(t')dt'$ oscillates between $0.5/\tau_m$ and $1.5/\tau_m$ with the same frequency as the flagella beat cycle $\Omega =2\pi\, 50\,\rm{Hz}$.
	The integrated friction resulting from Eq.~\eqref{eq_gammav_osc} is given by $G^{\rm{osc}}(t)=\frac{1}{\tau_m} + \frac{\sin (\Omega t)}{2\tau_m}$, whereas the integrated friction for $\Gamma_v(t)=2\delta(t)/\tau_m$ is given by $G(t)=\frac{1}{\tau_m}$.
	We choose the parameters of $\Gamma^{\rm{osc}}_v(t)$ such that the average integrated friction $G(t)$ is the same as for $\Gamma_v(t)=2\delta(t)/\tau_m$, while it oscillates at the flagella oscillation frequency.
	This allows us to evaluate the impact of the functional form of $\Gamma_v(t)$ on the extracted propulsion-force correlations based on a plausible model of flagella oscillations.
	Extracting the friction correlation under the assumption of $\Gamma_v(t)$ given by Eq.~\eqref{eq_gammav_osc} and using Eq.~(4) in the main text results in the blue line shown in Fig.~\ref{fig_gammar_diff_gammav}.
	The extracted correlation exhibits very similar behavior to the correlation extracted for $\Gamma_v(t)=2\delta(t)/\tau_m$, as shown in Fig.~\ref{fig_gammar_diff_gammav}.
	The ratio of force correlation magnitudes at time zero $\Gamma^{\rm{osc}}_R(0)/\Gamma^{\rm{del}}_R(0)\approx1.4$ showcases that the oscillating friction leads to similar but slightly higher estimates of the propulsion force.
	Here, $\Gamma^{\rm{osc}}_R(t)$ is the force correlation extracted under the assumption of oscillating $\Gamma^{\rm{osc}}_v(t)$, and $\Gamma^{\rm{del}}_R(t)$ is the correlation under the assumption of a pure delta contribution in $\Gamma_v(t)$, as in Eq.~(5) of the main text.
	Even though the magnitude of the extracted propulsion forces is influenced by the precise assumption of the flagella contribution to the hydrodynamic friction, the functional form of $\Gamma_R(t)$ is rather insensitive to the functional form of $\Gamma_v(t)$.
	
	In general, the shape of $\Gamma_R(t)$ is influenced by time-dependent friction $\Gamma_v(t)$.
	However, the most likely model for the flagella oscillation leading to an oscillatory $\Gamma^{\rm{osc}}_v(t)$ leads to almost the same functional form of $\Gamma_R(t)$ as for the frequency-independent $\Gamma_v(t)$ used for the extraction in the main text (Eq.~(5)), as shown in Fig.~\ref{fig_gammar_diff_gammav}.
	Further, the predicted values for the long-time diffusivity $D$, shown in the main text, remain valid independently of the exact functional form of $\Gamma_R(t)$.
	This is so because $C_{vv}(t)$ is a quantity directly extracted from the data and the functional forms of $\Gamma_R(t)$ and $\Gamma_v(t)$ have to reproduce the behavior of $C_{vv}(t)$.
	Thus, the integral over $C_{vv}(t)$ in Eq.~(10) of the main text, which yields the prediction for $D$, is independent of the data-driven model as long as $\Gamma_R(t)$ and $\Gamma_v(t)$ are self consistent.
	Nevertheless, the predicted $D$ depends on the localization noise that is estimated in our fitting procedure, described in section~\ref{sec_fitting}.
	Further, the estimated values for the power output of single cells is rather insensitive to the functional shape of $\Gamma_v(t)$.
	Here, mainly the total friction $G(t)$ determines the power output, which is expected to be in a similar order of magnitude when taking the flagella into account.

    \bibliographystyle{unsrt}
	\bibliography{nonEq.bib}